\documentclass[prl,superscriptaddress,twocolumn]{revtex4}
\usepackage{dcolumn}
\usepackage{graphicx}

\begin{document}

\title{Questioning the existence of a unique ground state structure
  for Si clusters}

\author{Waldemar Hellmann}
\affiliation{Institut f\"{u}r Physik, Universit\"{a}t Basel, CH-4056 Basel}

\author{R. G. Hennig}
\affiliation{Department of Materials Science and Engineering, Cornell
  University, Ithaca, NY 14853, USA }

\author{Stefan Goedecker}
\affiliation{Institut f\"{u}r Physik, Universit\"{a}t Basel, CH-4056 Basel}

\author{C. J. Umrigar}
\affiliation{Cornell Theory Center, Cornell University, Ithaca, New York 14853, USA}

\author{Bernard Delley}
\affiliation{Paul Scherrer Insitut, WHGA/123, CH-5232 Villigen PSI, Switzerland}

\author{T. Lenosky}
\affiliation{879 Meadowview Dr, Columbus Ohio 43224}

\begin{abstract}
  Density functional and quantum Monte Carlo calculations challenge
  the existence of a unique ground state structure for certain Si
  clusters. For Si clusters with more than a dozen atoms the lowest ten isomers
  are close in energy and for some clusters entropic effects can
  change the energetic ordering of the configurations. Isotope pure
  configurations with rotational symmetry and symmetric configurations
  containing one additional isotope are disfavored by these effects.
  Comparisons with experiment are thus difficult since a mixture of 
  configurations is to be expected at thermal equilibrium.
\end{abstract}


\maketitle

The determination of the structure of clusters is a difficult task.
The standard experimental techniques such as X-ray diffraction and NMR
methods that allow to determine the atomic positions in crystals and
molecules are not applicable to clusters~\cite{Hartke_1}.  The main
source of experimental information, ion mobility
measurements~\cite{Ion_Mob}, gives only crude information about the
overall shape of a cluster. The exact atomic positions of all the
atoms forming the cluster remain unknown.  For this reason
computational simulations provide a viable alternative to the
experimental approach, which has been widely used for silicon
clusters.  From the theoretical point of view the ground state
structure of a solid state system is determined by the global minimum
of the Born-Oppenheimer potential energy surface.  Finding the global
minimum requires global optimization algorithms.  Two problems arise
in this context. First, most global optimization algorithms give no
guarantee for finding the global minimum within a finite amount of
computer time.  Second, the Born-Oppenheimer energy surface has to be
calculated with very high precision.

Concerning the first point there is now a large amount of agreement
between different methods for medium size clusters containing up to 19
atoms~\cite{z_7_11, z_12_20}. Genetic algorithms~\cite{gen_Hartke_1,
Ho_Nature, Jackson_1}, the big-bang method~\cite{Jackson_2}, the basin
hopping method~\cite{Scheraga_BH, wales_BH, z_last} and the minima
hopping method~\cite{minhop} give typically similar or even identical
results. The discrepancies are rather due to different
exchange-correlation functionals in different
investigations~\cite{Yoo_Zeng_Si15_Si20}.

\begin{figure}[h]
  \begin{center}
    \setlength{\unitlength}{1cm}
    \begin{picture}(8.5,4.18)
      \put(-1.1,-0.9){\includegraphics{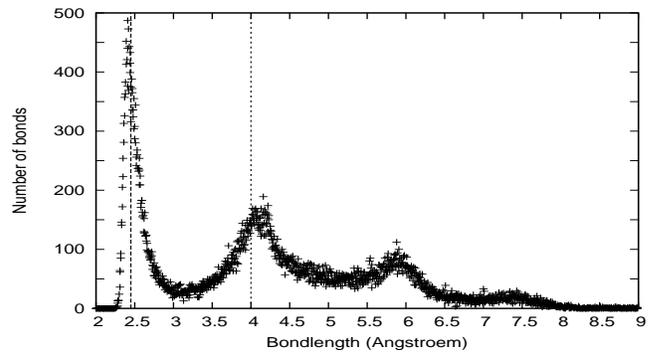}}
    \end{picture}
    \caption{ \label{bondlength} The bond-length distribution averaged
      over various low-lying Si17-configurations.  The two vertical
      lines indicate the 1st and the 2nd nearest neighbor distances in
      the crystal.}
  \end{center}
\end{figure}

The existence of a well defined ground state structure is generally
taken to be granted for silicon clusters. Silicon clusters are,
however, very different from bulk silicon where the second lowest
configuration (a fourfold coordinated defect~\cite{wooten_gap}) is 2.4
eV higher than the crystalline ground state. Clusters are frustrated
systems, where most of the atoms cannot adopt their favorite fourfold
coordination~\cite{Hartke_1}. This can lead to small energy
differences between different configurations. The significant
deviations of the clusters bond lengths from the crystalline bond
lengths shown in Fig.~\ref{bondlength} illustrate this frustration.

\begin{figure}[h]
  \begin{center}
    \setlength{\unitlength}{1cm}
    \begin{picture}(9.1,6.2)
      \put(-0.5,-0.9){\includegraphics{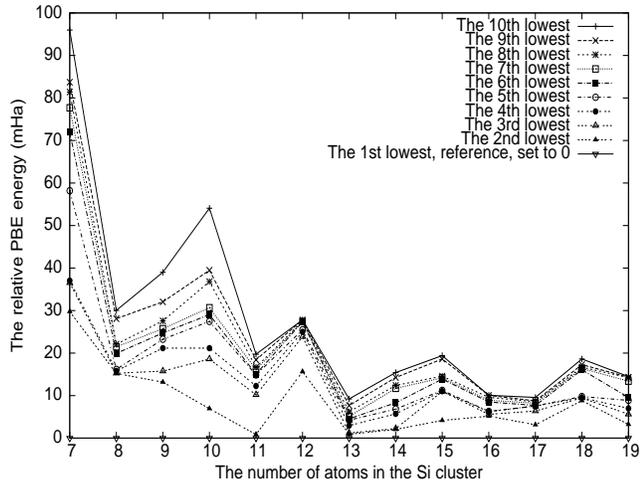}}
    \end{picture}
    \caption{ \label{clusters} The dependence of the PBE energy interval
      for the lowest 10 configurations on the cluster size.}
  \end{center}
\end{figure}

In this work we did not only search for the ground state
configurations of silicon clusters with up to 19 atoms, but for a
large number of low energy configurations.  This is possible with the
dual minima hopping method (DMHM)~\cite{DMHM}, which has the property
that it explores higher and higher energy configurations after having
found the global minimum. Fig.~\ref{clusters} shows the first major
result of our investigation, the energies of the 10 lowest
configurations of silicon clusters containing 7 to 19 atoms.  The
energy difference between the global minimum and the second lowest
minimum is 0.8~mHa for Si$_{11}$, 0.9~mHa for Si$_{13}$, 2.1~mHa for
Si$_{14}$, 3.1~mHa for Si$_{17}$ and 3.2~mHa for Si$_{19}$.  For 
Si$_{13}$ and Si$_{17}$ the 10 lowest configurations are in an
interval of roughly 10~mHa.  Since room temperature corresponds to
$\sim$1~mHa, entropic effects play an important role for these
clusters.

\begin{figure}[h]
\begin{center}
\setlength{\unitlength}{1cm}
\begin{picture}(20.0,13.0)

\put(-0.8,10.1){\includegraphics{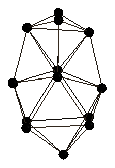}}
\put(0.7,10.75){Si$_{13}$}

\put(1.3,10.1){\includegraphics{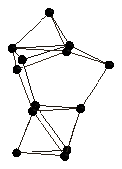}}
\put(2.65,10.75){Si$_{13a}$}

\put(3.5,10.1){\includegraphics{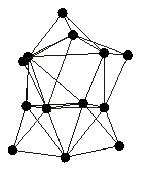}}
\put(4.8,10.75){Si$_{13b}$}

\put(5.45,10.1){\includegraphics{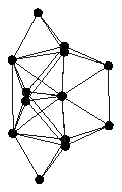}}
\put(6.85,10.69){Si$_{13c}$}

\put(-0.55,7.5){\includegraphics{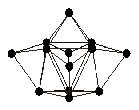}}
\put(0.68,8.0){Si$_{13d}$}

\put(1.44,7.5){\includegraphics{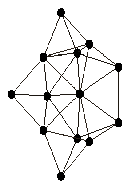}}
\put(2.67,8.0){Si$_{13e}$}

\put(3.55,7.5){\includegraphics{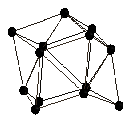}}
\put(4.85,8.0){Si$_{13f}$}

\put(5.59,7.46){\includegraphics{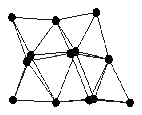}}
\put(6.82,8.0){Si$_{13g}$}

\put(-0.5,4.45){\includegraphics{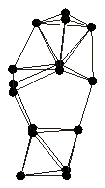}}
\put(0.8,4.85){Si$_{16}$}

\put(1.42,4.45){\includegraphics{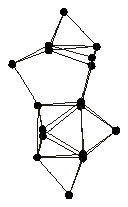}}
\put(2.94,4.85){Si$_{17}$}

\put(3.42,4.4){\includegraphics{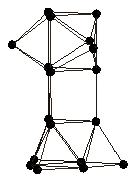}}
\put(4.98,4.85){Si$_{18}$}

\put(5.62,4.45){\includegraphics{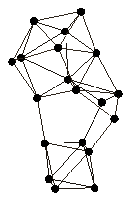}}
\put(7.26,4.85){Si$_{19}$}

\put(-0.5,1.54){\includegraphics{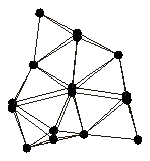}}
\put(1.0,2.41){Si$_{16a}$}

\put(2.3,1.66){\includegraphics{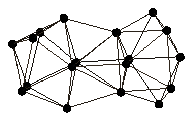}}
\put(3.8,2.53){Si$_{18a}$}

\put(5.3,1.65){\includegraphics{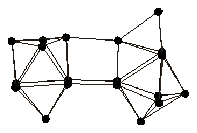}}
\put(6.7,2.50){Si$_{19a}$}

\put(-0.4,-0.81){\includegraphics{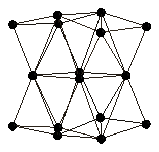}}
\put(1.0,-0.11){Si$_{16b}$}

\put(2.4,-0.70){\includegraphics{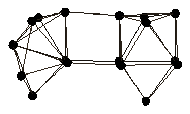}}
\put(3.76,-0.03){Si$_{17a}$}

\put(5.23,-0.71){\includegraphics{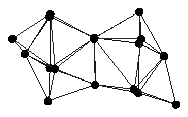}}
\put(6.6,-0.03){Si$_{17b}$}

\end{picture}
\caption{ \label{geometries} Symmetrized geometries of low-lying Si$_{13}$, Si$_{16}$, Si$_{17}$, Si$_{18}$ and Si$_{19}$ isomers.}
\end{center}
\end{figure}

The results of Fig.~\ref{clusters} were obtained with the
PBE~\cite{pbe} functional.  Even though this functional is considered
to be among the most accurate ones, its accuracy is clearly
insufficient to determine unambiguously the energetic ordering of the
configurations.  For this reason we have performed the most accurate
electronic structure calculations that are feasible for these systems,
namely quantum Monte Carlo (QMC) simulations.  The QMC calculations
are performed using the CHAMP code developed by Umrigar and Filippi.
The 1s, 2s and 2p electrons of Si are eliminated using a relativistic
Hartree-Fock pseudopotential~\cite{trail05}.  A Slater-Jastrow type
wave function is used as the trial wave function.  The orbitals of the
Slater determinant are taken from a DFT calculation with the
GAMESS~\cite{gamess} code using the B3LYP functional. The parameters
of the Jastrow function describing electron-electron, electron-nuclear
and electron-electron-nuclear correlations are optimized in
variational Monte Carlo using energy minimization~\cite{umrigar05}.
Diffusion Monte Carlo method calculates the final energies, which are
presented in Table~\ref{energies}. The corresponding configurations
are shown in Fig.~\ref{geometries}.  The QMC energies have error bars
of the order of 1~mHa which is just enough to discriminate between the
different energies. Even though the Monte Carlo results change the
energetic ordering of the PBE results, the central feature remains.
Different configurations have energies that are nearly identical.
Table~\ref{energies} also shows that the various high quality basis
sets used by different electronic structure programs give slightly
different answers that might change the energetic ordering.

The new low-lying structures Si$_{16a}$, Si$_{16b}$, Si$_{17a}$,
Si$_{17b}$, Si$_{18a}$ and Si$_{19a}$ which were found with DMHM and
the reference structures Si$_{16}$~\cite{z_last},
Si$_{17}$~\cite{Ho_Nature}, Si$_{18}$~\cite{Ho_Nature} and
Si$_{19}$~\cite{Jackson_1} were already presented in~\cite{DMHM}.  The
structure Si$_{13}$ was found by Ho {\it et al.}~\cite{Ho_Nature}, the
rotationally symmetric Si$_{13d}$ structure was recently proposed by
Hartke~\cite{Tek_Hart}, Si$_{13f}$ by R\"othlisberger {\it et
  al.}~\cite{rothl_13} and Si$_{13g}$ by Jeong {\it et
  al.}~\cite{Jeong}.  Using DMHM ~\cite{DMHM} we found new low-lying
structures Si$_{13a}$, Si$_{13b}$, Si$_{13c}$ and Si$_{13e}$.  From
the QMC results in Table~\ref{energies} we conclude that the
Si$_{13d}$ and Si$_{13a}$ configurations are the lowest energy
structures. The new Si$_{13a}$ structure found with DMHM contains the
stable Si$_{6}$ subunit~\cite{Si13_i}.

\begin{table}[t]
  \caption{The energy differences in mHa between
    the low energy geometries Si$_{13a}$, Si$_{13b}$,
    Si$_{13c}$, Si$_{13d}$, Si$_{13e}$, Si$_{13f}$, Si$_{16a}$,
    Si$_{16b}$, Si$_{17a}$, Si$_{17b}$, Si$_{18a}$, Si$_{19a}$
    and the reference structures Si$_{13}$, Si$_{16}$, Si$_{17}$,
    Si$_{18}$ and Si$_{19}$, proposed earlier as global minima
    in DFT.  The Gaussian~\cite{gaussian} calculations used the
    6-311G(2d) basis and the  DMol3 2005~\cite{DMOL_1, DMOL_2}
    calculations the extended basis set. The CPMD~\cite{cpmd}
    calculations were performed with an accurate
    pseudopotential~\cite{pseudo} with a 35~Ry plane
    wave cutoff and a 30~\AA\ simulation cell.}
  \label{energies}
  \begin{ruledtabular}
    \begin{tabular}{l d d d d d}
      Cluster & \multicolumn{1}{c}{B3LYP} & \multicolumn{1}{c}{PBE} &
      \multicolumn{1}{c}{PBE} & \multicolumn{1}{c}{PBE} &
      \multicolumn{1}{r}{DMC} \\
      & \multicolumn{1}{c}{GAUS.} & \multicolumn{1}{c}{DMOL } &
      \multicolumn{1}{c}{CPMD} & \multicolumn{1}{c}{GAUS.} \\
      \hline
      Si$_{13}$  &   0.0 &   0.0 &  0.0 &  0.0 &   0.0 \\
      Si$_{13a}$ & -11.7 &   6.3 &  7.1 &  5.3 &  -3.3\pm1.0 \\
      Si$_{13b}$ &  -1.4 &   9.5 & 13.2 & 11.5 &   5.7\pm1.0 \\
      Si$_{13c}$ &   8.6 &   4.7 &  3.1 &  3.1 &   6.4\pm1.0 \\
      Si$_{13d}$ &  -6.7 &  -2.9 &  0.9 &  0.6 &  -5.7\pm1.1 \\
      Si$_{13e}$ &   3.6 &   3.9 &  4.1 &  4.2 &   6.1\pm1.0 \\
      Si$_{13f}$ &   4.2 &   7.0 &  8.5 &  8.0 &  10.6\pm1.0 \\
      Si$_{13g}$ &  -6.3 &  -0.8 &  2.6 &  2.3 &   0.0\pm1.1 \\
      Si$_{16} $ &   0.0 &   0.0 &  0.0 &  0.0 &   0.0 \\
      Si$_{16a}$ &  24.0 &  -0.2 & -6.0 & -3.7 &   9.9\pm1.4 \\
      Si$_{16b}$ &   9.5 &   0.9 &  0.9 &  1.4 &   8.3\pm1.4 \\
      Si$_{17} $ &   0.0 &   0.0 &  0.0 &  0.0 &   0.0 \\
      Si$_{17a}$ &   7.4 &   8.4 &  2.6 &  2.7 &   6.3\pm1.5 \\
      Si$_{17b}$ &  10.2 &  11.6 &  8.4 &  8.4 &  13.2\pm1.7 \\
      Si$_{18} $ &   0.0 &   0.0 &  0.0 &  0.0 &   0.0 \\
      Si$_{18a}$ &  31.3 &  17.1 & 10.0 & 10.2 &  23.9\pm1.6 \\
      Si$_{19} $ &   0.0 &   0.0 &  0.0 &  0.0 &   0.0 \\
      Si$_{19a}$ &  15.4 &  -1.0 & -4.2 & -1.9 &   2.1\pm1.7 \\
    \end{tabular}
  \end{ruledtabular}
\end{table}

After having discussed the limitations of computational approaches in
determining the total energy of silicon clusters with the necessary
accuracy let us discuss the physical effects that can change the
energetic ordering. For the Si$_{13}$, Si$_{13a}$ and Si$_{13d}$
clusters we have zero point energies of 24.0, 25.0 and 24.5~mHa. For
the Si$_{19}$ and Si$_{19a}$ we have 38.8 and 38.0~mHa. So the
differences of the zero point energies are all of the order of mHa and
thus not negligible, but do not change the energetic ordering for the
clusters we studied.  In order to study the entropic effects we
calculated the the rotational and vibrational free energy based on the
harmonic frequencies obtained from density functional (PBE)
calculations~\cite{textbook}. The translational free energy does not
depend on the configuration and was therefore not considered. If one
compares the sum of the rotational and vibrational free energy for
non-symmetric configurations, one typically finds differences of about
0.5~mHa at room temperature and about 1~mHa close to the melting point
of the clusters~\cite{REV_MT}.  This might change the energetic
ordering, but we did not find a case where it actually does.

\begin{figure}[t]
  \begin{center}
    \setlength{\unitlength}{1cm}
    \begin{picture}(8.78,4.8)
      \put(-0.82,-1.04){\includegraphics{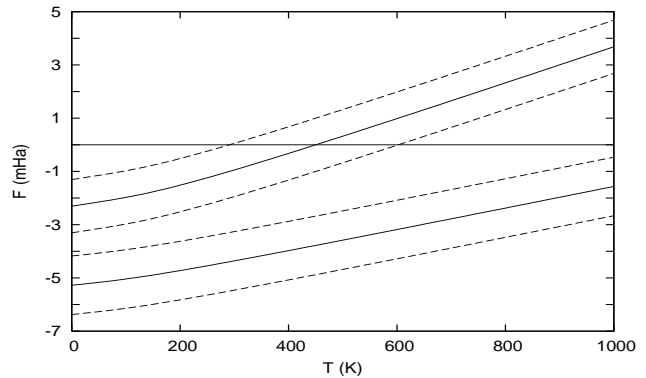}}
    \end{picture}
    \caption{ \label{freeen} The sum of the electronic (QMC, with errors),
      rotational and vibrational (including zero point) free energy
      contributions for Si$_{13a}$ (upper band) and Si$_{13d}$ (lower band) 
      configurations as a function of temperature with the
      Si$_{13}$ free energy chosen as reference energy 
      (solid line, set to 0).}
\end{center}
\end{figure}

The situation is different if one compares a symmetric with a
non-symmetric configuration.  Silicon occurs in nature mainly as
$^{28}$Si or $^{29}$Si isotope.  The predominant isotope for silicon
$^{28}$Si (abundance $\sim$92\%~\cite{isotope}) has mass 28 and no
nuclear spin, the $^{29}$Si isotope (abundance
$\sim$5\%~\cite{isotope}) has mass 29 and nuclear spin 1/2. When
studying configurations with rotational symmetry, we will consider
pure clusters consisting only of $^{28}$Si atoms since the presence of
a $^{29}$Si atom would destroy the rotational symmetry. One can easily
estimate from the abundancies of the isotopes that $\sim$34\% of
Si$_{13}$ clusters will be pure clusters.  For such a cluster with
rotational symmetry, the order of the rotational subgroup enters into
the formula for the rotational free energy.  This leads to a weaker
decrease of the free energy for symmetric configurations compared to
non-symmetric configurations and thus favors non-symmetric structures.
In Fig.~\ref{freeen} we present the free energy curves for the
structures Si$_{13a}$ and Si$_{13d}$ as a function of temperature with
the Si$_{13}$ free energy chosen as reference energy.  The width of
the bands for Si$_{13a}$ and Si$_{13d}$ represents the statistical
errors in the QMC energies with respect to that of structure
Si$_{13}$.  For the symmetric Si$_{13a}$ configuration the order of
the rotational subgroup is 3, for Si$_{13d}$ it is 2 and for Si$_{13}$
it is 1.  This leads to a reversal of the energetic ordering of the
structures Si$_{13}$ and Si$_{13a}$ in the interval between 250 and
650~K.  Because of the entropic effect the Si$_{13}$ configuration,
which is the highest at zero temperature, becomes the lowest at
temperatures above 1000~K.  At room temperature the Si$_{13a}$ and
Si$_{13d}$ bands are separated by an energy gap in the range between
$\sim$1.2mHa and $\sim$5.5~mHa.  This corresponds to a Boltzmann
weight in the range between 0.7\% and 30\%.

These considerations are only valid for clusters consisting purely of
$^{28}$Si atoms.  The presence of a $^{29}$Si isotope destroys the
rotational symmetry.  One can estimate from the abundancies of the
isotopes that $\sim$24\% of Si$_{13}$ clusters will contain one
$^{29}$Si isotope.  If one $^{28}$Si atom with nuclear spin 0 is
replaced by a $^{29}$Si isotope which has spin 1/2, the nuclear
wave-function is a doublet and additional degeneracy comes from the
fact that the isotope can replace any of the atoms.  For a
non-symmetric cluster with $N$ atoms the degeneracy is thus $2 N$. For
a symmetric cluster that has several equivalent atoms the degeneracy
is however reduced.  In the case of the Si$_{13a}$ structure there are
for instance only 5 non-equivalent sites, Si$_{13d}$ has 6 and
Si$_{13}$ has 9.  The nuclear entropy thus favors Si$_{13}$ over
Si$_{13d}$ by $ - k T \ln(\frac{9}{6})$ which is $\sim$ 0.4~mHa at
room temperature. In addition, the vibrational and rotational entropy
contributions are slightly changed by the presence of an isotope
leading to an effect of the same order of magnitude.

\begin{figure}[t]
  \begin{center}
    \setlength{\unitlength}{1cm}
    \begin{picture}(8.78,6.1)
      \put(-0.68,-1.1){\includegraphics{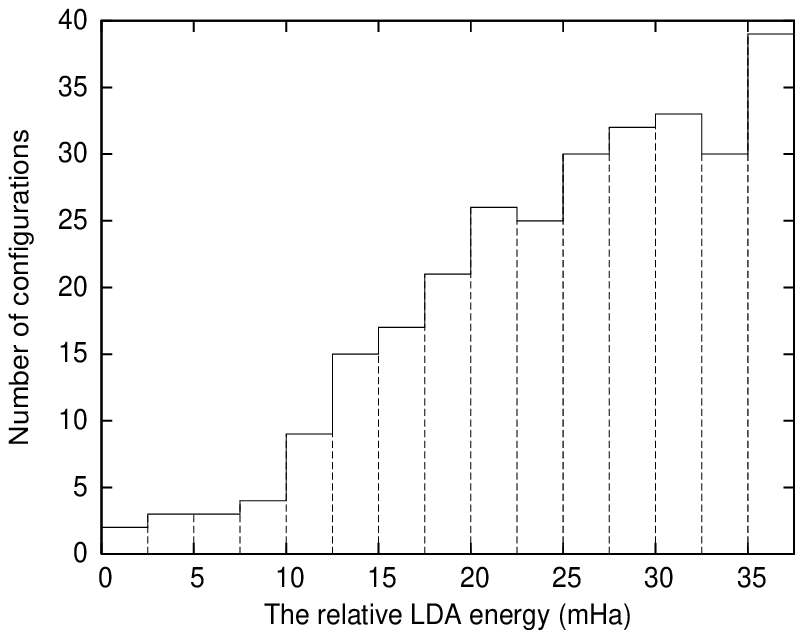}}
      \put(-0.182,2.492){\includegraphics{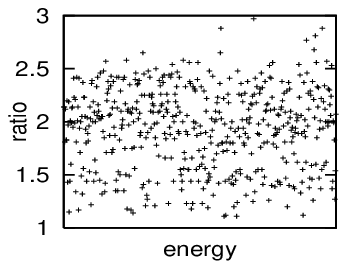}}
    \end{picture}
    \caption{ \label{density} The configurational density of states for
      Si$_{17}$. The inset shows the ratio of the largest and smallest 
      eigenvalues of the moment of inertia tensor for various low-lying
      Si$_{17}$ configurations.}
  \end{center}
\end{figure}

Up to now we have concentrated on the 10 lowest structures.
Considering higher lying configurations, the energetic spacing between
configurations decreases even further.  This can be inferred from the
fact the the configurational density of states, defined as the number
of configurations per energy interval, increases strongly.  This is
shown in Fig.~\ref{density} for the Si$_{17}$ cluster.

Several simulations have shown that the lowest energy structures for Si 
clusters with less than 20 atoms are non-spherical whereas larger clusters prefer
to be spherical~\cite{Hartke_1, Jackson_2}. In these studies the
non-spherical to spherical transition was obtained by considering 
the putative ground state configurations.  The inset in
Fig.~\ref{density} shows the ratio between the largest and smallest
eigenvalues of the moment of inertia tensor for various low-lying
Si$_{17}$ clusters. A ratio of 1 corresponds to a spherical geometry,
while larger values correspond to non-spherical structures.
For a given cluster size we observe in Fig.~\ref{density} that the 
occurrence of non-spherical and spherical structures is independent of
energy.

In summary, we have shown that there exists a large number of configurations for certain 
silicon clusters that are energetically extremely close to the ground state. This 
feature was observed for Si$_{13}$ and Si$_{19}$ and it will presumably 
be even more important for larger cluster sizes that were not studied in this work. 
As a consequence, entropy effects that are usually neglected 
can change the energetic ordering of the lowest configurations. Entropy
disfavors symmetric clusters Si$_n$ in the range $13\le n \le 19$
which contain in most cases no $^{29}$Si isotope or one $^{29}$Si
isotope. Larger clusters will on average contain more than one
$^{29}$Si isotope and the symmetry related effects discussed above do
not exist.  However, for larger clusters the 10 lowest configurations
can be expected to lie within an even smaller interval. The entropic
effects not related to symmetry considerations might thus easily
change the energy order of clusters with more than 19 atoms.  Even if
there is no reordering, different structures can be so close in free
energy that a mixture of two or more configurations will be found at 
thermal equilibrium.  As a consequence measured properties of clusters
can be some average of the properties of
several low-lying isomers.\\
We thank Markus Meuwly for discussions, X. C. Zeng and A. Tekin for
the Si$_{13}$ cluster data, the Swiss National Science Foundation for
financial support and the staff of the computing center at the
University of Basel for technical support. The work at Cornell
University was supported by NSF grant EAR-0530301. Computational
resources were provided by OSC, NERSC and NCSA.

\end{document}